\documentclass[a4paper,11pt]{article}
\usepackage{jinstpub}  

\title{(PREPRINT) Low-cost open-source high-frequency portable pulse counter for Raspberry Pi and its application to Xray transmission rate measurement}

\author{Aydin Tarik Zengin}

\affiliation{Istanbul S. Zaim University, Turkey}

\emailAdd{tarik.zengin@izu.edu.tr, tarik.zengin@gmail.com\}}

\abstract{Affordable electronics for instrumentation play a vital role in academia since research budgets are tight nowadays. 
In this paper a low-cost open-source high-frequency portable Raspberry Pi-based pulse counter  is presented.
Although it is designed as a $ 2 $ channel counter, it can be easily modified to a $ 4 $ channel counter. 
It provides a relay control interface for experimental-setup integration and counts pulses up to $ 10 $ MHz frequency.
Presented system is proved to be accurate by an Xray transmission rate measurement experiment.}


\begin{document}
\maketitle
\flushbottom

\section{Introduction}
\label{sec:intro}

Low-cost electronics are vital for researchers due to limited research budgets. 
Limited sources of open-source hardware electronics has barred research since some researchers cannot either afford or easily acquire the required laboratory electronic equipment.  
Recently there have been some efforts to provide affordable electronics that perform various kinds of measurements.
S. Ritt et. al. introduced an open-source software interface for their GHz rate waveform digitizer \cite{Ritt2010}.
A. E. Beltran developed open-source hardware, low-cost acquisition board with ADCs and various filters \cite{Beltran2013}.
E. Zahedi proposed open-source hardware and software dual-channel biosignal recorder \cite{Zahedi2013}.
K. Jin and Y. Song developed an open-source hardware, beaglebone-black-based sensor monitoring hardware \cite{Jin2016}.
G. Real et. al. combined the industrial controllers and data acquisition features in one open-source hardware and software board \cite{Real2018}.
As seen in the examples above, more and more open-source hardware and software designs were being introduced for the researchers.
This paper presents an affordable, portable, and open-source-hardware (OSHW) and open-source-software (OSSW) frequency/pulse counter for high frequency signals.
The hardware was designed as a Raspberry-Pi hat that can be directly connected to the I/O ports.
The software was implemented in Python on a Qt based graphical user interface running on Raspberry Pi.

\section{Design Methods}
In order to make the system affordable, it was designed as an extension board for Raspberry Pi (RPi).
Owing to its being a widely known and commonly used embedded system, the Rpi-based designs can be easily obtained and installed by scientists. 
Presented system consists of 2 components; the hardware and software parts.

\begin{figure}[!hbtp]
\caption{Counters' I/O connections}
\centering
\includegraphics[width=.95\linewidth]{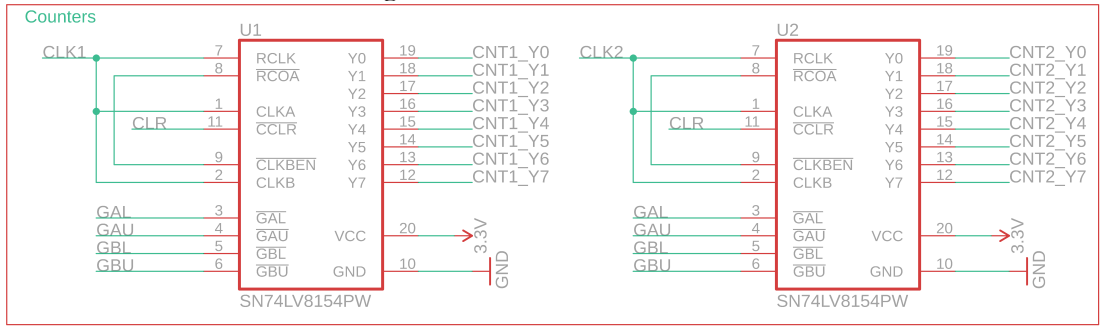}
\label{counter_conn}
\end{figure}

\subsection{Hardware Design}

\begin{figure}[htbp]
\caption{RPi-Counter Connections}
\centering
\includegraphics[width=0.6\linewidth]{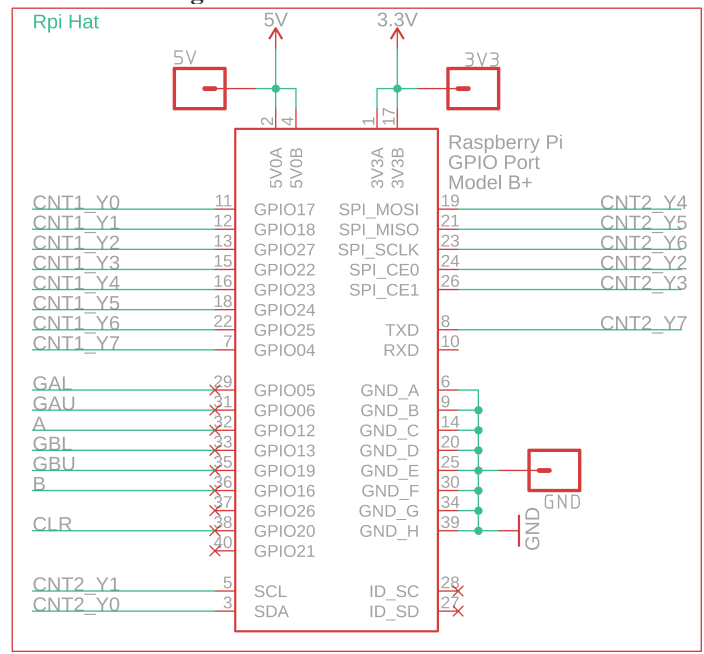}
\label{rpi_conn}
\end{figure}

Hardware system includes 2 independent 32-bit hardware counters (SN74LV8154)\cite{texas_counter} each has 1 \emph{clock} input, 1 \emph{clear} input, 4 \emph{control} inputs and 8 data outputs as shown in Figure.~\ref{counter_conn}.
Therefore most of the RPi I/O pins were used by the counters in this project.
32-bit output data were multiplexed to the output pins in bytes by using 4 control inputs GAL, GAU, GBL, GBU, ordered from least significant byte to most significant byte, respectively.
Both of the counters were connected to common control inputs.
Therefore the selected byte of the data was always the same for both counters.
Although the presented system was designed as a 2-channel 32-bit counter, it can easily be modified to function as a 4-channel 16-bit counter by disconnecting $\overline{RCOA}$ - $\overline{CLKBEN}$ and $CLKB$ - $CLKA$ connections.
In such case, the 3rd and 4th clock signals should be applied to $CLKB$ pins.
The pinout diagram of RPi showing where the counter pins are connected is given in Figure.~\ref{rpi_conn}.

\begin{figure}[hbtp]
\caption{Counter data structure}
\centering
\includegraphics[width=.6\linewidth]{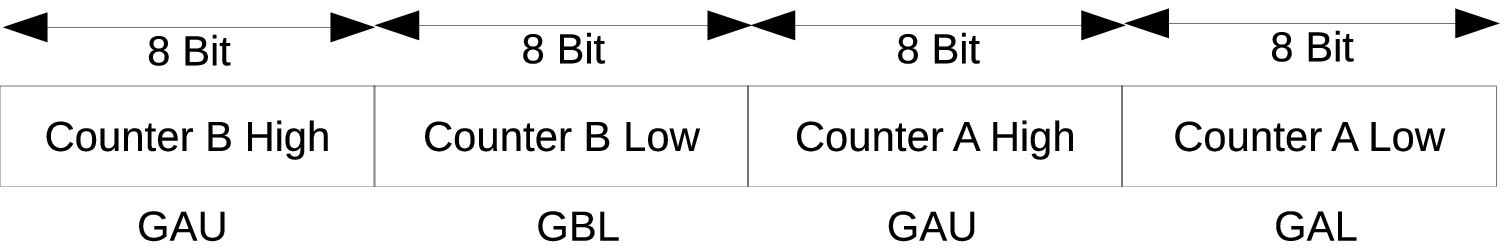}
\label{counter_data_structure}
\end{figure}

To acquire 32-bit data from the counter, each 8-bit section of it should be taken sequentially by iterating 4 control inputs as it's explained above.
Then a simple logic/arithmetic operation was applied to reconstruct the 32-bit data.
Data structure is shown in Figure.~\ref{counter_data_structure}.

\begin{figure}[htb]
\caption{Input Comparators}
\centering
\includegraphics[width=\linewidth]{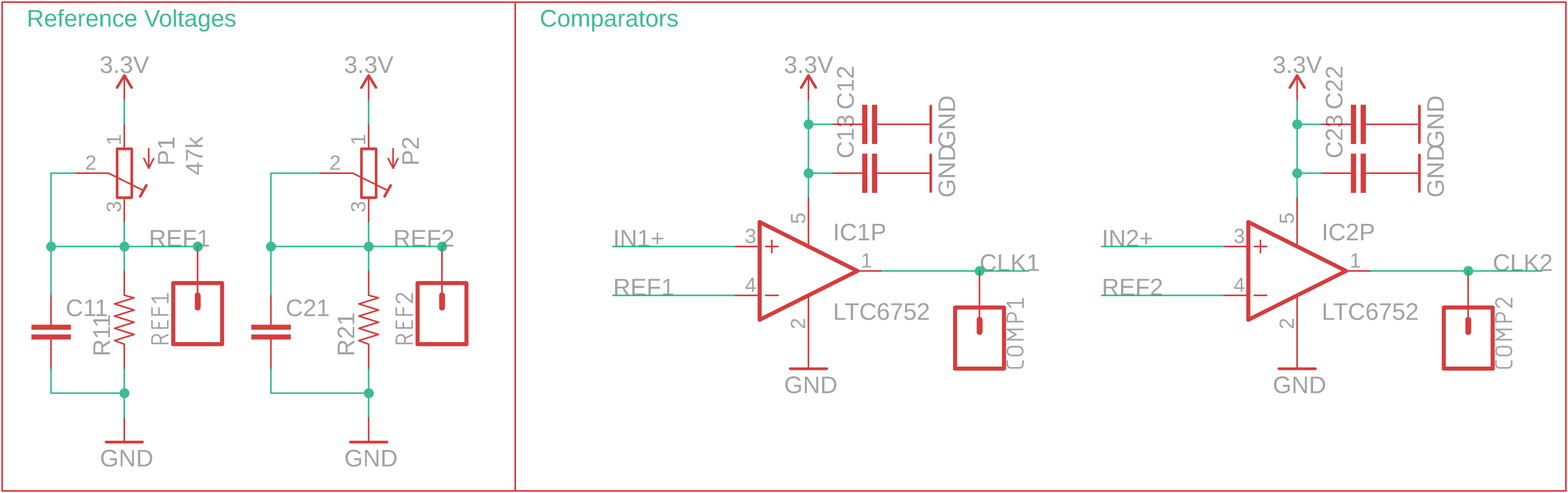}
\label{comparator}
\end{figure}

Each counter had an opamp comparator as their clock input as shown in Figure.~\ref{comparator}.
Comparators had a voltage divider at the inverting inputs to adjust the reference comparison voltage.
In this specific application, voltage divider resistors are chosen as $ R_{11} = R_{21} = 470 \Omega $ and $ P_1  = P_2 = 47k\Omega$.
Values can be adjusted according to the application.

\begin{figure}[!hbtp]
\caption{PCB Top View}
\centering
\includegraphics[width=0.7\linewidth]{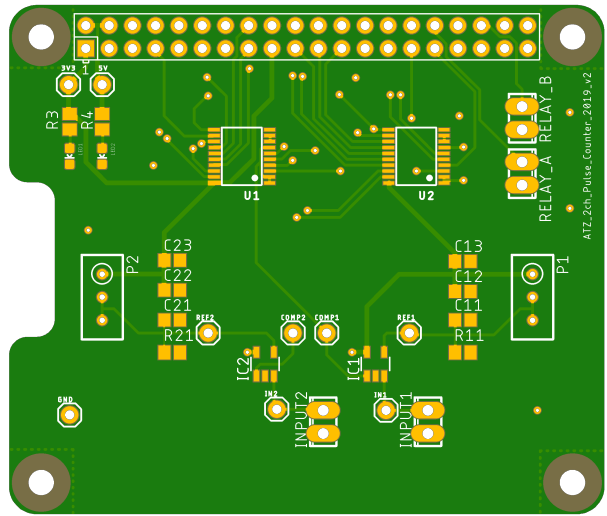}
\label{pcb}
\end{figure}

Printed Circuit Board (PCB) was designed by using Eagle EDA software.
Signal inputs and relay outputs were placed on the sides of the board for easy accessibility.
Voltage divider potentiometers were chosen as top-adjust worm-gear trimpots for easy operation.
Test pins for \emph{reference voltages}, \emph{comparator outputs} and \emph{source voltages} were placed around the board for easy troubleshooting.
The design is shown in Figure.~\ref{pcb}.
Schematics and board design files were published\footnote{https://github.com/atzengin/RPi-Frequency-Counter} freely under the GPL license. 
It can freely be changed, improved and republished.

\subsection{Software Design}
Graphical user interface (GUI) was programmed by using Python and Qt library.
Data acquired from the board were shown in the middle part of the screen with timestamps for each interval set by the operator.

Hardware counters count automatically on reception of a trigger signal fired by the opamp comparators.
Interval of count retrieval can be set in milliseconds from the GUI.
On every interval set by the GUI, data were multiplexed and combined by the software.

\begin{figure}[hbtp]
\caption{Graphical User Interface}
\label{gui1}
\centering
\includegraphics[width=.8\linewidth]{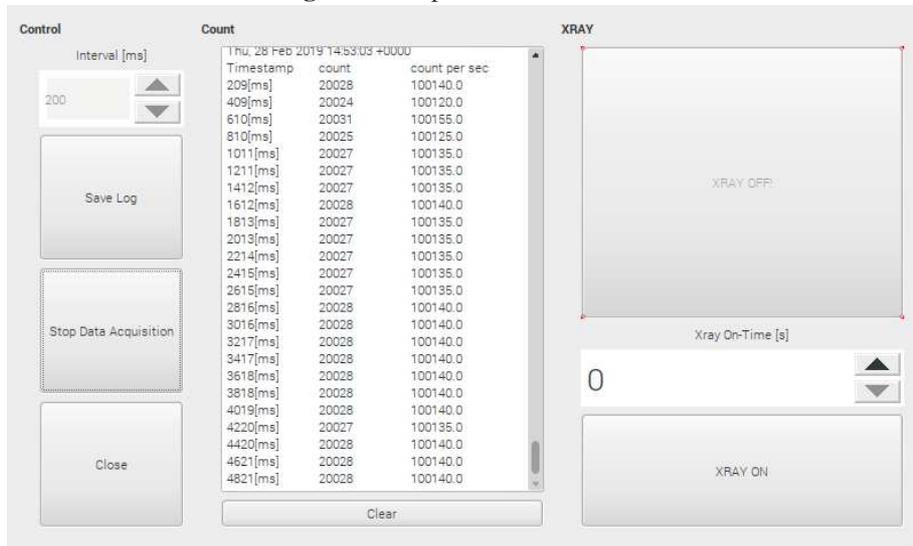}
\end{figure}

The GUI provides an on-off interface for relays to control any hardware in the experimental setup.
In this specific example, an X-ray source was controlled via \emph{XRAY ON} and \emph{XRAY OFF} buttons.
It was also possible to set an on-time in seconds for the relay that was on for the chosen time and turned off automatically at the end of this period.
\emph{Save Log} feature saves the count logs with timestamps as a .txt file.
The source code was published\footnote{https://github.com/atzengin/RPi-Frequency-Counter-Software} freely on Github under the GPL license.
Touchscreen- and mouse-operation-compatible GUI is shown in Figure.~\ref{gui1}.

\begin{figure}[hbtp]
\caption{Final design with touchscreen and I/O ports}
\centering
\includegraphics[width=.8\linewidth]{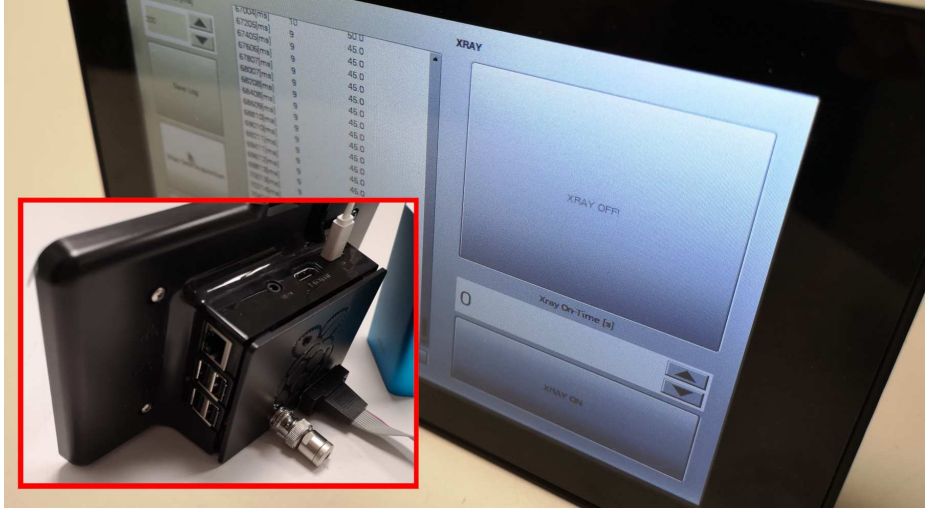}
\label{screen}
\end{figure}

The final design combining the hardware and software is shown in Figure.~\ref{screen}.
Back side of the casing having the signal input (only 1 channel was implemented for this specific application) and relay connector can be seen in the lower left frame.
It can run on battery power source and does not require keyboard/mouse, hence, making it a portable system.

\section{Accuracy Test}
\begin{figure}[!hbtp]
    \centering
    \begin{minipage}{.5\textwidth}
        \centering
        \includegraphics[width=\linewidth, height=0.3\textheight]{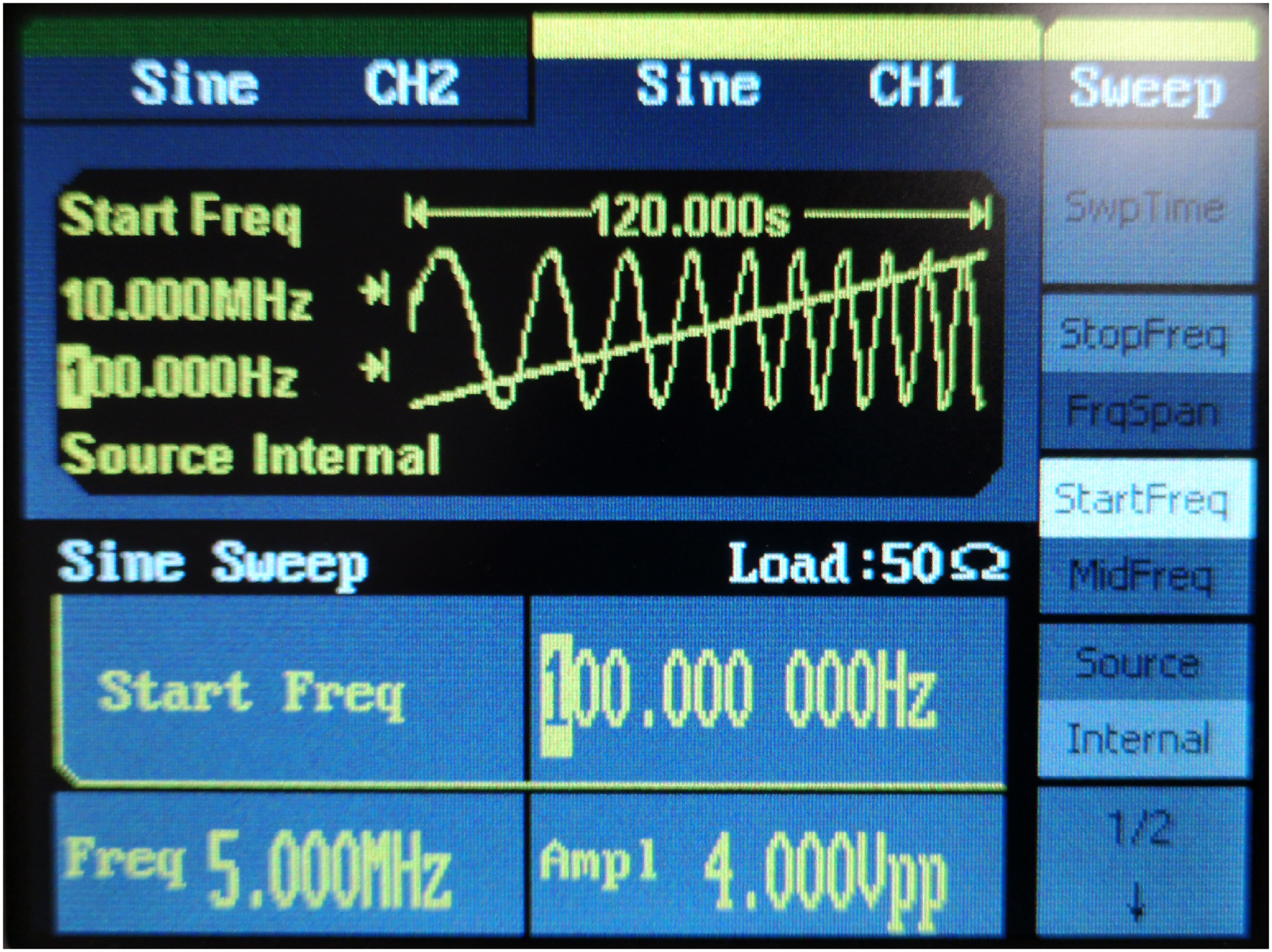}
        \caption{Sweep settings}
        \label{sweep_settings}
    \end{minipage}%
    \begin{minipage}{0.6\textwidth}
        \centering
        \includegraphics[width=\linewidth, height=0.3\textheight]{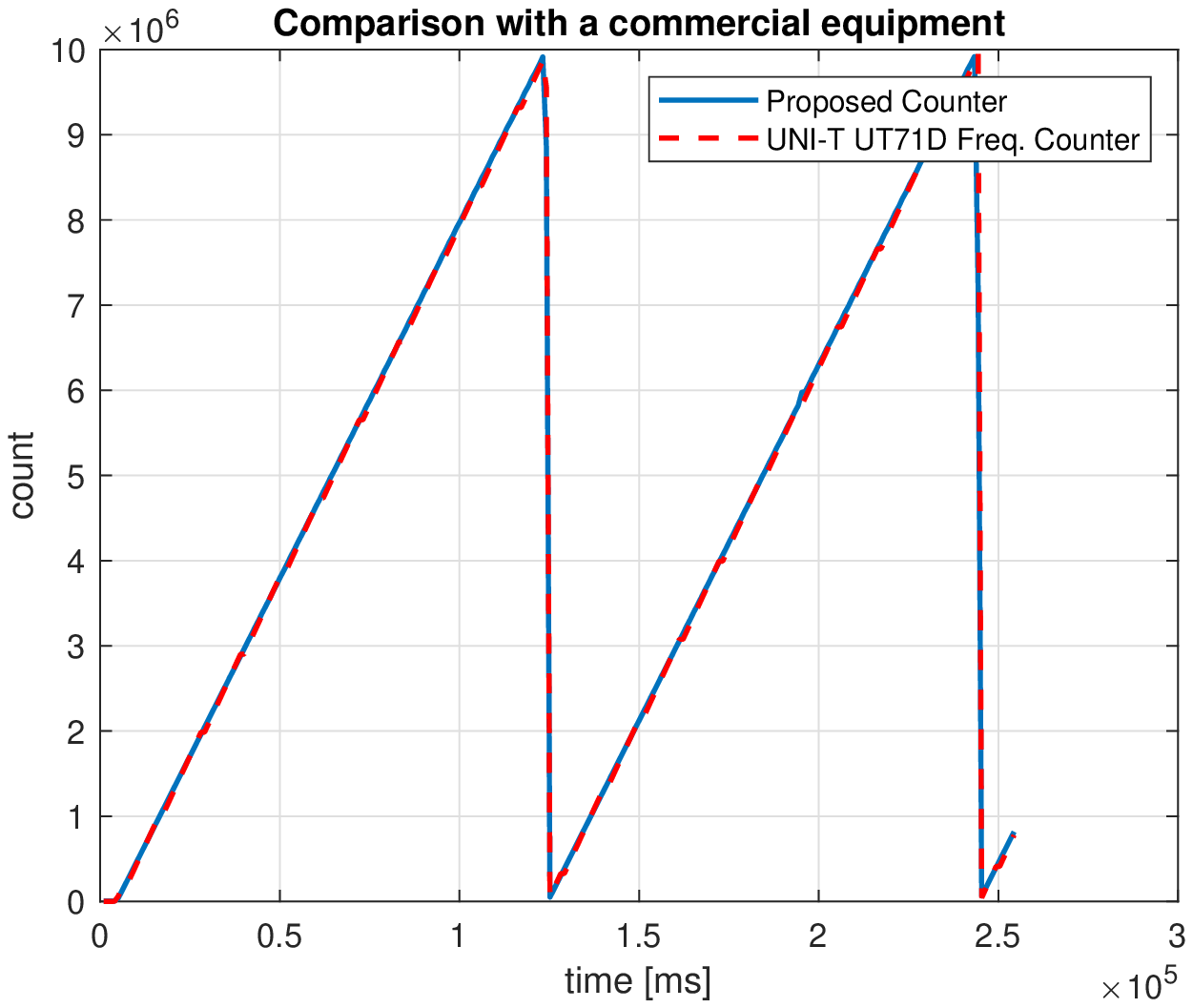}
        \caption{Comparison of proposed counter}
        \label{sweep_comp}
    \end{minipage}
\end{figure}

In order to gauge the system's accuracy, an AC signal sweeping from $100$Hz to $10$MHz was applied to the counter.
Even though the proposed design supported millisecond level intervals, it was set to $ 1000 $ms from the GUI since the frequency-counter-to-be-compared (UNI-T UT71D) supported this interval only.
The count was displayed in real time during the sweep and the logs saved to a .txt data file.
Sweep signal settings and count graph visualization of sweep data are given in Figure.~\ref{sweep_settings} and Figure.~\ref{sweep_comp}, respectively.
The graph shows that the system perfectly reflected the linearly incremented frequency over time.
Therefore the system was sufficiently accurate.


\section{Application of Transmission Rate Measurement}
\begin{figure}[!hbtp]
\caption{Experimental Setup}
\label{exp_setup}
\centering
\includegraphics[width=\linewidth]{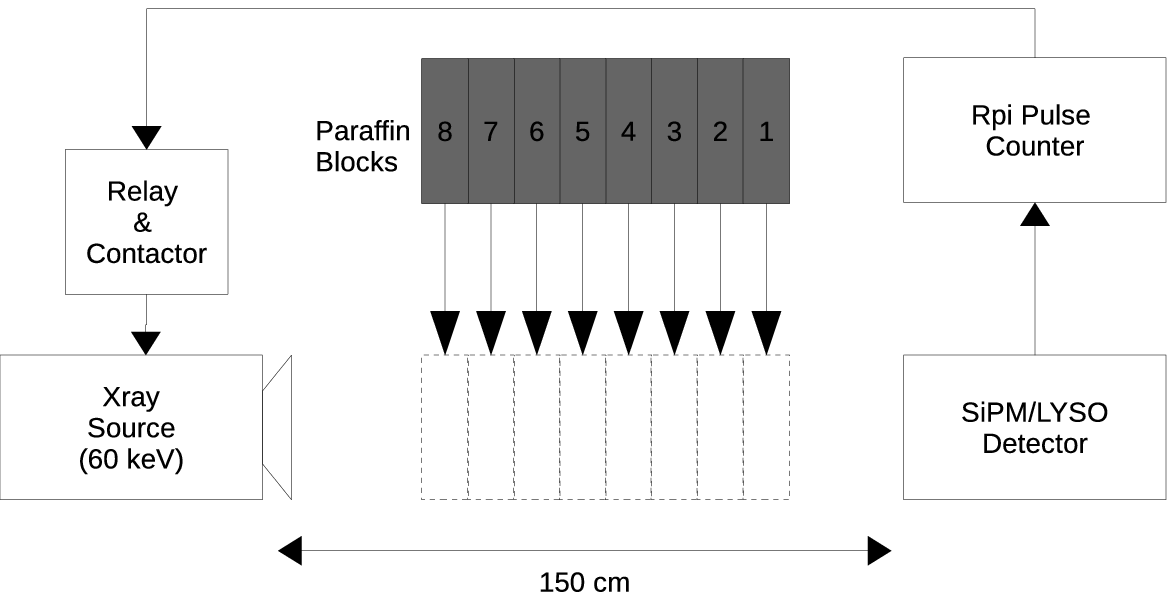}
\end{figure}

In order to show the accuracy and to clarify the usage of proposed design, an experiment was conducted.
Setup is given in Figure.~\ref{exp_setup}.
A $60$ keV Xray source and an SiPM (Silicon Photo Multiplier) detector were laid out in a straight line with 150 cm distance between them.
Detector has a LYSO(Ce) - Lutetium Yttrium Orthosilicate ($Lu_{1.8}Y_{.2}SiO_5:Ce$) scintillator for catching photons.
Proposed counter received the output signal of detector and counted the pulses which fired by detected photons \citep{Iren2016} \citep{Yetkin2017}.
The pulse count gave the intensity of photons reached the detector.
When there was no paraffin blocks in between the Xray source and detector, the count was recorded as $I_0$ (initial intensity of photons).

As measuring the transmission rate by using the proposed counter was the goal of the experiment, paraffin blocks -each is $14$ mm thick- were placed in between the Xray source and detector.
Intensity of photons is given as $ I = I_0 e^{-\mu x}$ where $\mu$ is the linear attenuation coefficient ($\mu_{paraffin}=0.18 cm^{-1}$ ) and $x$ is the distance traveled in the material.
Number of paraffin blocks placed in the setup changed intensity $I$ in connection with $x$.

\begin{figure}[!hbtp]
\caption{Transmission rate of 60 keV photons through paraffin}
\label{transmission}
\centering
\includegraphics[width=\linewidth]{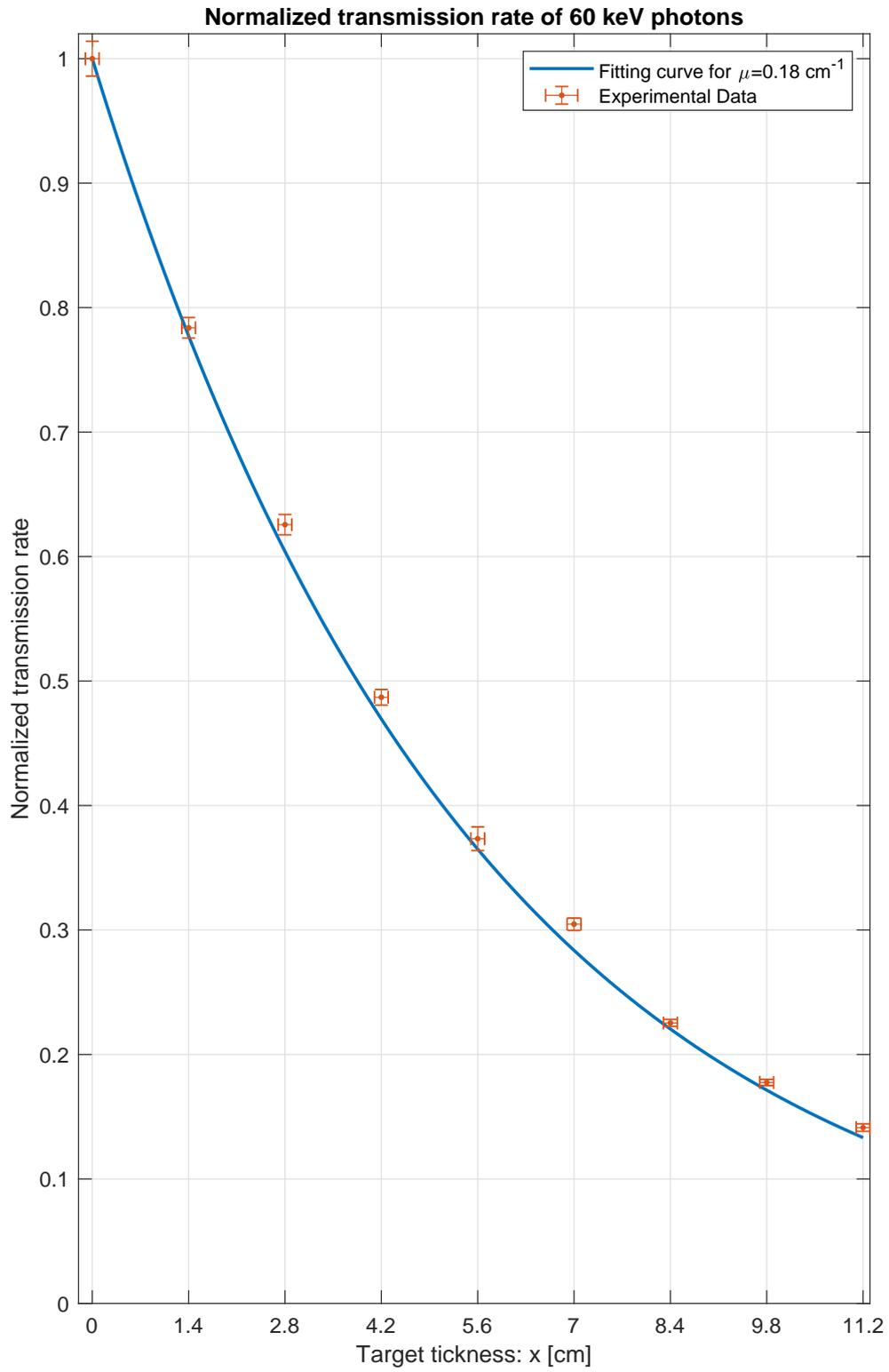}
\end{figure}

Transmission rate for $60$ keV photons through paraffin was calculated using the formula given above for different number of blocks were placed in front of the detector.
Comparison of the calculated and measured transmission rate is given in Figure.~\ref{transmission}.
Xray source was controlled by the proposed counter to shoot for a specific period of time that was set in the GUI.
In this experiment, shooting time was selected as $5$s.
Standard deviation of transmission during the Xray shoot and measurement error of the paraffin blocks' thickness ($\pm 1mm$) were shown as the vertical and horizontal error bars, respectively, in the figure.
It clearly showed the proposed counter was accurate.

\section{Conclusion}
In this paper, a low-cost frequency/pulse counter is presented.
It was consisted of 2 32-bit independent channels with adjustable threshold levels.
The design was published freely for both the hardware and software on Github.
It can freely be downloaded and assembled by anyone who needs a low-cost pulse counting solution.
The transmission rate test showed that system was accurate enough for the mid and high frequency pulses.
Digitally adjustable threshold levels and increasing the channel number were left as the future works.

\acknowledgments

I would like to thank T. Yetkin and F. Ozok from Mimar Sinan Fine Arts University, and N. Erduran and G. Erdemir from Istanbul S. Zaim University for their valuable insights and comments.

\bibliographystyle{JHEP}
\bibliography{tarik_paper}

\end{document}